%% file: ecrts_rt-cloud_2025.tex
\documentclass[conference]{IEEEtran}
\IEEEoverridecommandlockouts
\usepackage{cite}
\usepackage{amsmath,amssymb,amsfonts}
\usepackage{algorithmic}
\usepackage{graphicx}
\usepackage{textcomp}
\usepackage{xcolor}
\def\BibTeX{{\rm B\kern-.05em{\sc i\kern-.025em b}\kern-.08em
    T\kern-.1667em\lower.7ex\hbox{E}\kern-.125emX}}

\usepackage{array}
\newcolumntype{C}[1]{>{\centering\arraybackslash}p{#1}}
\newcolumntype{M}[1]{>{\centering\arraybackslash}m{#1}}

\ifCLASSOPTIONcompsoc
\usepackage[caption=false,font=normalsize,labelfont=sf,textfont=sf]{subfig}
\else
\usepackage[caption=false,font=footnotesize]{subfig}
\fi

\begin{document}

\title{Offloading tracing for real-time systems using a scalable cloud infrastructure\\
\thanks{This Project is supported by the Federal Ministry for Economic Affairs and Climate Action (BMWK) on the basis of a decision by the German Bundestag.}
}

\author{
\IEEEauthorblockN{Schmidt, David Jannis}
\IEEEauthorblockA{\textit{Institute of Technology and Computer Science} \\
\textit{University of Applied Science}\\
Gießen, Germany \\
david.schmidt-3@mni.thm.de}
\and
\IEEEauthorblockN{Fridman, Grigory}
\IEEEauthorblockA{\textit{} \\
\textit{Schmidt Embedded Systems GmbH}\\
Gießen, Germany\\
grigory.fridman@schmidt-embedded-systems.de}
\and
\IEEEauthorblockN{von Zabiensky, Florian}
\IEEEauthorblockA{\textit{Institute of Technology and Computer Science} \\
\textit{University of Applied Science}\\
Gießen, Germany \\
florian.von.zabiensky@mni.thm.de}
}

\maketitle

\graphicspath{{./imgs/}}


\begin{abstract}

Real-time embedded systems require precise timing and fault detection to ensure correct behavior. 
Traditional tracing tools often rely on local desktops with limited processing and storage capabilities, which hampers large-scale analysis.
This paper presents a scalable, cloud-based architecture for software tracing in real-time systems based on microservices and edge computing. 
Our approach shifts the trace processing workload from the developer’s machine to the cloud, using a dedicated tracing component that captures trace data and forwards it to a scalable backend via WebSockets and Apache Kafka. 
This enables long-term monitoring and collaborative analysis of target executions, e.g., to detect and investigate sporadic errors.
We demonstrate how this architecture supports scalable analysis of parallel tracing sessions and lays the foundation for future integration of rule-based testing and runtime verification. 
The evaluation results show that the architecture can handle many parallel tracing sessions efficiently, although the per-session throughput decreases slightly as the system load increases, while the overall throughput increases.
Although the design includes a dedicated tracer for analysis during development, this approach is not limited to such setups. 
Target systems with network connectivity can stream reduced trace data directly, enabling runtime monitoring in the field.

\end{abstract}


\section{Introduction}\label{sec:introduction}

Real-time applications in embedded systems have become an integral part of daily life. They are designed to perform specific tasks within strict time constraints. Failure to comply can have a wide range of consequences depending on the situation and product, ranging from annoyed users and financial losses to life-threatening situations. Therefore, the analysis of such systems and the rapid detection of errors through debugging and testing are crucial to ensure their correct behavior and reliability \cite{gliwa_embedded_2021}. Usual interactive debugging techniques, such as breakpoints, stepping, and manual memory monitoring, offer a way to get basic insights into the execution of such applications. However, they do not provide any information about the timing behavior of the system. \emph{Tracing} describes an alternative technique that counteracts this problem. 
We define a trace as an independent stream of events that enables reconstruction, visualization, and analysis of system behavior by including timing parameters and execution information.
In embedded systems, \emph{hardware-based tracing} records events at the machine instruction level using dedicated hardware, logic, and memory on the chip \cite{arm_arm_2023}. In contrast, \emph{software-based tracing} logs events in standard RAM and relies on instrumentation within the application software. Due to the high memory and CPU overhead of instruction-level tracing, software-based methods typically operate at the function level. In real-time operating systems, instrumentation of task state changes and interrupt service routine entry or exit allows for detailed timing analysis at the OS level. This facilitates the rapid identification of both bottleneck causes and potential optimization points for runtime performance.\\
Various commercial and open-source tools support hardware- and software-based tracing for a wide range of processors and microcontrollers \cite{percepio_tracealyzer_2025}, \cite{segger_microcontroller_gmbh_segger_2025}. These tools typically run as local desktop applications on the development computer and require a direct serial connection to the target device. High data volumes from tracing events or complex analyses can strain system resources, leading to performance issues or data loss due to the limited memory and computing power of the development machine. Offloading tracing to the cloud addresses these limitations by taking advantage of virtually unlimited processing, storage, and communication resources. This aligns with the potential for long-term and collaborative analyses, as well as runtime verification.

This paper presents a cloud-based approach for software-based tracing of real-time embedded systems, focusing on a highly scalable architecture using microservices. The potential architecture for offloading tracing data processing to the cloud will be examined in more detail. In order to reduce complexity and data load, the focus is limited to software-based tracing at the operating system and application level. To leverage cloud scalability and efficiently handle high data traffic, the proposed solution must support dynamic scaling. This raises the following research questions:

\begin{itemize} 
    \item How can a cloud-based, scalable architecture for tracing in real-time systems be designed to enable efficient analysis and fault diagnosis of high-density systems?
    \item How can the scalable operation of such a tool in the cloud be ensured?
    \item How does the number of concurrent tracing sessions affect processing time and overall throughput?
\end{itemize}

The paper is structured as follows: Section \ref{sec:related_work} provides an overview of the related work. In Section \ref{sec:architecture}, we present our suggestion for our architecture, which we explain in more detail in Section \ref{sec:implementation}. Section \ref{sec:evaluation} shows the evaluation of the architecture before we come to a conclusion in Section \ref{sec:conclusion} and outlines future work.

\section{Related Work}\label{sec:related_work}

\subsection{Tracing  for embedded systems}

Software-based tracing uses software instrumentation to generate tracing events. This has an impact on system performance, which can be significant depending on the level of detail \cite{convent_hardware-based_2018}. There are various commercial systems that can examine the traces of embedded systems. Many of them use a developer computer to store these traces, which excludes collaborative analysis by developers and limits trace capacity \cite{percepio_tracealyzer_2025}, \cite{segger_microcontroller_gmbh_segger_2025}. A commercial tool uses a cloud system and therefore relies on a similar architecture as we present in this paper, but it turns out that a PC is used between the target device and the cloud \cite{bonnevier_continuous_2025}. We, on the other hand, rely on smaller but still powerful tracer hardware, which delivers more flexibility and smaller costs.

Hardware-based tracing, on the other hand, provides non-intrusive raw data and does not influence the system. However, this means that many traces have to be collected and stored \cite{convent_hardware-based_2018}. This is where the strengths of a cloud-based solution become apparent, as a scalable number of resources are available for storing the tracks.

The work of Pagano et al. shows an infrastructure for trace analysis and management of traces \cite{pagano_soc-trace_2012}, \cite{pagano_generoso_trace_2013}. This system is also aimed at embedded systems and offers some parallels to our architecture. In our architecture, however, we rely on powerful and easily scalable microservices and time series databases that are suitable for traces, rather than relational databases such as those used by Pagano et al.

\subsection{Scalable cloud architectures}

In the context of Linux environments, there are numerous tracing projects in existence.
It is important to note the importance of the Linux Tracing Toolkit next generation (LTTng) in this context \cite{lttng_lttng_2025}.
LTTng is an open-source Linux tracing framework that is designed to facilitate the analysis of system execution by means of tracing formats or tools for recording and subsequent analysis.
As demonstrated by the barectf project, the tracing is not confined to Linux systems \cite{barectf_barectf_2025}.
In the context of the LTTng ecosystem, TraceCompass is utilized as a tool for visual analysis \cite{eclipse_foundation_trace_2025}. 
Its functionality is analogous to that of tools such as SEGGER SystemView.
An examination of the TraceCompass reveals the existence of an additional initiative that targets modern, scalable C/C++ development workflows. 

We use microservices as the underlying architecture. Microservices have become popular in recent years and bring far-reaching advantages in terms of system performance in contrast to monolithic systems \cite{pahl_microservices_2016}, \cite{mohottige_microservices-based_2024}. This is an important characteristic, particularly in terms of scalability. Studies have shown that microservice-based applications perform better than monolithic implementations when it comes to memory and CPU usage, as well as response time \cite{dheeraj_konidena_securely_2024}. Other advantages include the abstraction of individual subtasks of the overall system and the resulting improved maintainability.

There are a large number of commercial and open source systems for distributed tracing, such as Jaeger and Zipkin in conjunction with OpenTelemetry, to name only a few \cite{jaeger_tracing_jaeger_2025}, \cite{zipkin_zipkin_2025}, \cite{opentelemetry_opentelemetry_2025}. Scaling for a large number of clients is possible, and these frameworks often also use a microservice-based architecture. These have established themselves in recent years, particularly in the area of web development, whereas there is no support for the embedded domain.
For frameworks such as Datadog and NewRelic, on the other hand, there is support for IoT devices, but the focus is on monitoring and is not dedicated to tracing embedded real-time systems \cite{datadog_datadog_2025}, \cite{new_relic_inc_new_2025}.

To summarize, it can be said that the tools mentioned do offer possibilities for distributed tracing, but the use case to date has focused on cloud or web software. In this work, we try to use the advantages of these frameworks and combine them with embedded real-time systems.

\section{Architecture of the cloud-based tracing system}\label{sec:architecture}

In this Section we introduce the cloud-based system for tracing embedded targets and going deeper into detail.
Based on the research questions in Section \ref{sec:introduction}, the architecture should address the following problems:
\begin{itemize}
    \item Processing and aggregation of real-time traces recorded from an embedded system.
    \item Compression and transmission of traces over a network connection.
    \item Concentration of computing and storage power in the cloud.
\end{itemize}

\begin{figure}[ht]
    \centering
    \includegraphics[width=0.489\textwidth]{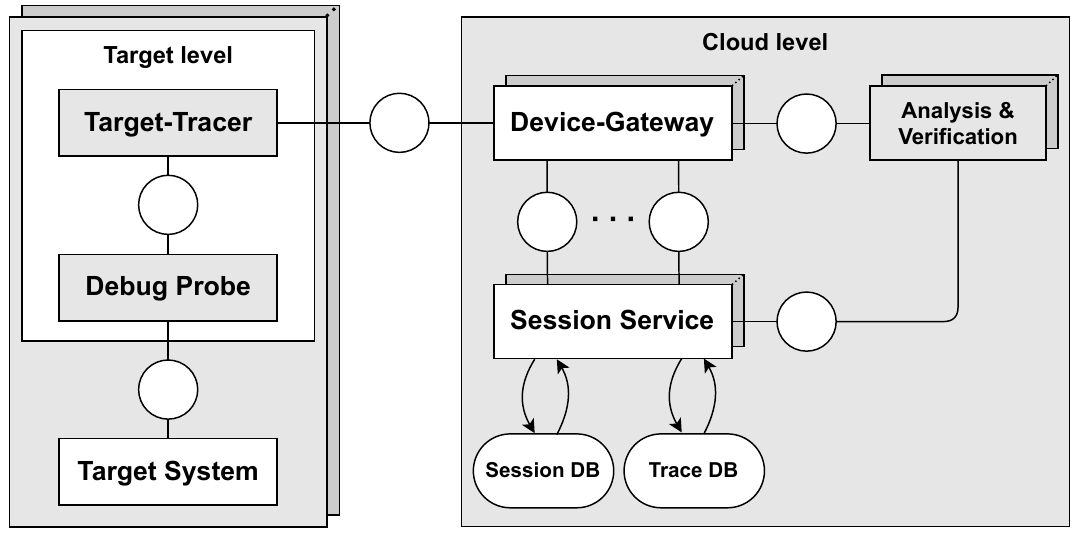}
    \caption{The architecture of the cloud-based embedded tracing system as a FMC-Diagram.}
    \label{fig:architecture}
\end{figure}

Our architecture consists of two levels: A \textit{target level} and a \textit{cloud level} as shown in Figure \ref{fig:architecture}. A third component, the \textit{target system}, is the embedded system to be analyzed, which provides real-time traces that are read out by a debug probe.
The target level focuses mainly on the aggregation and efficient transfer of traces to the cloud. In this way, we can outsource them to the cloud and thus scale better for a large number of target systems. The cloud level provides the computing and storage capacity and contains the actual application logic. The main task is to persist and manage traces and to provide a session service with which tracing can be started or stopped on a target. 

\subsection{Target Tracer}\label{subsec:target_tracer}

The target level comprises both the target system to be analyzed and an additional component: the \textit{target tracer.} It offers sufficient computing power and memory capacity and accordingly consists of its own computing unit and a debugger. In this way, we can further abstract the target system by outsourcing computationally intensive operations that do not have to be performed by the target system. In addition, target systems without a network connection or low-performance systems can be operated in this way. 
The debugger represents the connection to the embedded target and ensures that the traces generated by the target are transferred to the target tracer.
Incoming traces are aggregated and compressed on the target tracer computing unit implemented in a single software service,  ensuring more efficient transmission to the cloud.

It should be noted that this abstraction level enables the use of non-intrusive tracing methods, as the target system does not have to use software instrumentation, thus generating software traces. Accordingly, the debugger can also be used for non-intrusive tracing methods. The architecture presented here provides a separate target tracer for each connected target system.
Furthermore, the architecture allows for the use of a target tracer to provide traces. It is also possible to send traces directly to the cloud structure via a network interface.

\subsection{Device Gateway}\label{subsec:device_gateway}
The device gateway is part of the cloud level and provides a distribution service for other cloud components. By a bidirectional network connection, information is transferred between the cloud and the device.
The designated service has the task of forwarding all requests from the cloud to the target tracer and distributing incoming traces from the target tracer into the cloud services. Requests from the cloud come from users who want to start or stop a trace on a target.
Incoming traces and messages from the target tracer are distributed via a total of five specific publish-subscribe channels. 
Those communication channels are: (i) A \textit{request channel} and (ii) a corresponding \textit{response channel} for communication commands between services. These are used to start and stop tracing sessions on a target, as well as to receive acknowledgments of commands. (iii) A channel for \textit{connection events} received when a new target tracer connects to the cloud. Also, (iv) an \textit{interrupt channel} where information about trace interruptions and errors is transmitted. There is also a separate channel for (v) \textit{tracing events} from the target. 
Remote services in the cloud can subscribe to these channels.

The basic idea of microservices lies in the independence and statelessness in relation to messages from individual, stand-alone applications. In this way, individual services can be replaced on the fly.
We use precisely this statelessness in our cloud infrastructure by outsourcing the transmission of messages. The messages are saved and stored in an external database and distributed via the various channels. In this way, we pave the way for the use of microservices.

\subsection{Session Service}\label{subsec:session_service}
The subscribers of all five communication channels are in a service we call session service. 
It has the task of managing a \textit{tracing session} and receiving tracing events from the device gateway. We define a tracing session as an independent execution on a target device that provides a specific trace consisting of a sequence of temporal events. Two databases are connected to this service. One of these databases is used to persist all metadata from a tracing session. The other database stores all incoming traces. Scalability is to be achieved by providing this service as a microservice.
The service also provides session information and controls other services. It is the API gateway to interact with the system. 

\subsection{Trace analysis and verification}\label{subsec:trace_analysis}
Another component of the architecture is trace analysis and verification. Connecting this component to the cloud makes it possible to analyze incoming trace events. 
This also enables collaborative analysis for developers, which other tools in the embedded systems domain do not offer. 
Incoming traces can be tested and verified using rule-based checks. In theory, this enables runtime verification by integrating frameworks such as the TeSSLa ecosystem \cite{kallwies_tessla_2022}.

Real-time requirements can arise with this component if it is used for the runtime verification of a system. Tracing Events are generated on the target, e.g. to check system states, then transmitted to the cloud and verified there. The verification result is sent back to the target, where it is switched to a safe state, or the operation is continued based on the test results. In this way, runtime verification is outsourced from the target system to the cloud. The execution of the runtime verification must take place within defined latencies, which are specified by the real-time requirements of the target.

We would like to mention at this point that the architecture offers this possibility, but the implementation we show in the next chapter does not include these components but is planned for future work.

\begin{figure*}[htbp]
\centering
\includegraphics[width=\textwidth]{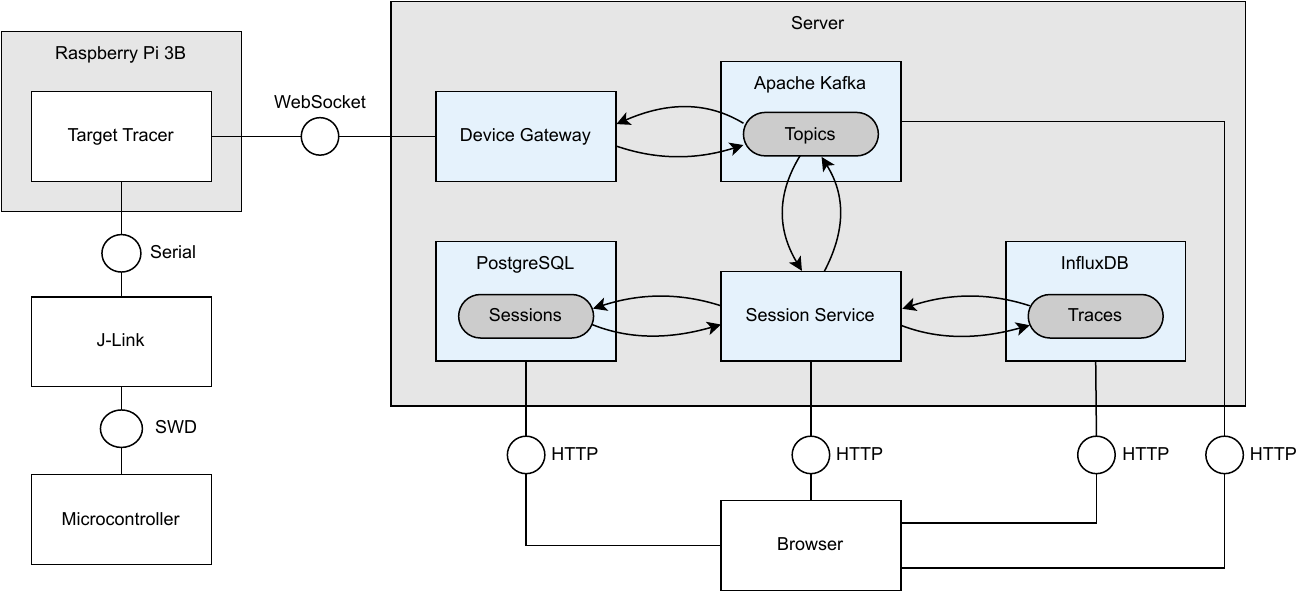}
\caption{Containerised test system of our architecture using docker.}
\label{fig:deployed_architecture}
\end{figure*}

\section{Implementation}\label{sec:implementation}
In the following section we show the implementation of our cloud-based tracing system. 

\subsection{Target system}\label{subsec:target}
The aim of the target is to generate events at runtime, which are then transferred to the target tracer.
We use an ARM Cortex processor for the implementation, which has a JTAG and SWD debug interface, so that we can exchange data via this interface. We use SEGGER Real Time Transfer (RTT), a lightweight byte-based protocol that can be used to transfer data bidirectionally with a transfer rate of up to 2~MB/s \cite{segger_microcontroller_gmbh_real_2021}. This means that we transfer tracing events to the debugger and commands from the debugger to the controller.
RTT uses the targets' memory to temporarily store generated traces at runtime, which are then transferred over the debugger to the Target Tracer application. We use SEGGER SystemView to generate tracing events on our target \cite{segger_microcontroller_gmbh_segger_2025}.

\subsection{Target Tracer Implementation}\label{subsec:target_tracer_impl}
As the implementation of the target tracer depends on the debugger as well as the protocol used for data transmission of events or commands, this is specified by the target. This means that in our Target Tracer application we need a receiving point for the debugger data stream and the RTT protocol support. On the other hand, our application must be able to send commands to the target. The target tracer application sends a message to the cloud at startup to signal that a new Target Tracer is connected. This is used for activation in the cloud. Only after activation can a user send commands to this Target Tracer and thus start tracing on a connected target.

We use WebSockets in conjunction with Protocol Buffer (Protobuf) messages for all bidirectional communication between the Target Tracer and the device gateway \cite{google_llc_protocol_2025}.
With Protobuf messages, we have, on the one hand, readable and easily expandable messages during implementation, as well as efficient transmission during runtime compared to other well-known data formats, such as JSON, due to a smaller data footprint \cite{popic_performance_2016}. 

In addition, we have developed a \textit{file~mode}. With this mode, we achieve a simulation of a target. This allows us to better investigate the scalability of the implemented architecture without having to connect a large number of real targets and target tracers.
If the target tracer is used in file mode, a pre-recorded trace file is loaded instead of a connection via a debugger to a real target.
The target tracer then reads the entire pre-recorded trace file and saves it in a memory buffer.
When a command to start a tracing session is sent from the cloud to the target tracer, all events stored in the buffer are sent sequentially at the highest possible frequency determined by the receiving capacity of the device gateway.

\begin{table*}[htbp]
    \footnotesize
    \caption{The system configuration and processing time results.}
    \centering
    \begin{tabular}{|c|C{1.8cm}|C{1.3cm}|C{1.8cm}|C{1.9cm}|}
        \hline
         Configuration no. & Session Service instances & Number of sessions & Throughput per session & Overall throughput \\
         \hline
         1&  1& 1 & 13,558\,events/s & 13,558\,events/s\\ 
         \hline
         2&  1& 50 & 879\,events/s & 43,966\,events/s\\ 
         \hline
         3&  3& 1 & 14,421\,events/s & 14,421\,events/s\\ 
         \hline
         4&  3& 50 & 1,911\,events/s & 95,560\,events/s\\ 
         \hline
         5&  5& 1 & 13,616\,events/s & 13,616\,events/s\\ 
         \hline
         6&  5& 50 & 2,569\,events/s & 128,445\,events/s\\ 
         \hline
    \end{tabular}

    \label{tab:evaluation_scaling_conf}
\end{table*}

\subsection{Device Gateway Implementation}\label{subsec:dev_gateway_impl}
For communication between microservices in the cloud, we have chosen a combination of asynchronous request-response and event patterns for data distribution to increase performance. We refer to these as messages and use the publish-subscribe pattern to distribute these messages.
We use Apache Kafka as the central distribution unit in our architecture \cite{sax_apache_2018}. Kafka offers low latency, error tolerance, and the ability to transport large amounts of data using a publish-subscribe message broker system.  It is also scalable, as the number of message brokers can be adjusted. We use a total of five Kafka topics, which were introduced in Section \ref{subsec:device_gateway}. 
The application consists of three components, which are subdivided as follows, whereby each component is implemented as a single task:
\begin{itemize}
    \item The first component provides WebSocket servers, which are required for the connection of the target tracers. These connections must also be managed. A new WebSocket server is provided for each connection to a target tracer. In this way, we aim to achieve better isolation of the connections, although the overhead increases and load balancing is not optimal. All messages sent to a target tracer are distributed to the two other components of the application.
    \item The second component forwards incoming requests that are received from the Kafka topic. It is therefore the only component that receives messages from a Kafka topic. The messages are used for control purposes, e.g., to start and stop tracing sessions, and are sent by a user. 
    \item The third component in turn forwards all incoming tracing events of a single target tracer to Apache Kafka. In this component, we have also placed our decoder for transforming a received tracing event into a protobuf message. This component sends messages to the corresponding Kafka topics for incoming responses, connection, interrupt, and tracing events that are sent by a target system.
\end{itemize}

\subsection{Session Service Implementation}\label{subsec:session_service_impl}
The session service, in turn, represents the remote station for messages sent by the device gateway, but also publishes messages or control commands in the direction of the device gateway. 
As this service is also responsible for persisting and providing the databases, two databases are linked to the service.
We use PostgreSQL as the database system for our session management and InfluxDB for storing the traces.
PostgreSQL is an efficient relational database for managing sessions. InfluxDB, on the other hand, is better suited as a time series database for storing traces, as it is specialized in storing time series data. This enables more efficient and higher-performance storage of this data compared to relational database models or other time series databases like TimescaleDB \cite{shah_performance_2022}.

We have divided the application logic of the microservice into two tasks. The first task performs our session service management. An additional API allows users to start tracing sessions and send control commands to a specific target tracer through the device gateway. This API can also be used to send queries to the session database.
The second task is responsible for processing the tracing messages sent by the device gateway via Apache Kafka. These are received in the task and stored in the InfluxDB or PostgreSQL database, depending on the topic. We use a state machine for each session to monitor the state, react to errors, and send different commands to a target tracer depending on the state.  The state of a session is stored in the database accordingly.
Our microservice operates stateless in this way, as all session-related information is stored in the database rather than in the service itself. This decouples the service from a session and thus ensures better scaling. Each service can therefore process a session-specific message when it is received, regardless of the session itself. If a service crashes, it can be replaced directly by another service without losing session-relevant information. This is ensured by extensive error handling. In the event of an error, docker restarts the crashed service.

\section{Evaluation}\label{sec:evaluation}
In this chapter we present an evaluation based on the implementation described in chapter \ref{sec:implementation}.

\subsection{Preliminaries}\label{subsec:preliminaries}
We use a SEGGER J-Link debugger in combination with a Rasberry Pi 3 Model B as a target tracer. This combination provides sufficient performance for the implementation of the target tracer. The target system consists of a microcontroller board equipped with a Cortex-M3 processor. We have connected the microcontroller board via the SWD debug interface. We collected various metrics and instrumented and monitored an application that generates our traces. 
The application executed on our microcontroller target generates approximately $3,200$ events per second. 

As mentioned in previous chapters, it is possible to simulate targets by replaying events from a pre-recorded file from the target tracer. Consequently, we have created a recording file for the microcontroller application which we use to evaluate the scaling. The recording file contains almost $31,726$ events at about 150 kB and lasts $10.012$ seconds.
The file-mode of the target tracer is particularly suitable for evaluating scaling. Traces read from the pre-recorded file are sent to the cloud immediately after conversion into protobuf events. Real targets may generate events with a small time delay between them. These are not taken into account in the file mode. In our case, we therefore achieve a higher, almost maximum throughput when evaluation with the file mode for the target tracer than when using real targets on which the described target application is running.

We use Docker for containerization to provide the entire cloud layer. Consequently, containers are provided for all the services described in the previous chapters. Additional Docker images are used for the databases, Apache Kafka, and for monitoring. This allows us to monitor all the implemented services as well as the message traffic between the services. We access the monitoring tools through a web interface. The structure of our system can be seen in Figure \ref{fig:deployed_architecture}. We have highlighted all Docker services in blue.

\subsection{Scaling}\label{subsec:scaling}
In order to evaluate the scaling of the presented architecture, we use the Docker Swarm mode. This feature creates a group of physical or virtual machines (swarm) in which Docker-based applications can run in a distributed manner. To collect the series of measurements for the evaluation, we used a PC with an Intel Core i9-9880H (8C/16T @ 2.3 GHz) and 32 GB RAM running Ubuntu 22.04 LTS. The databases use the local NVMe SSD to store persistent data.

The event throughput of one tracing session (\textit{throughput per session}) and all tracing sessions that run in parallel (\textit{overall throughput}) were evaluated. 
We define event throughput as the number of events that can be processed and stored in a given time. The duration of a tracing session, called the \textit{processing time}, was also examined. The processing time indicates how long it takes for a tracing session to be processed by the Session Services, from the first event (start tracing) to the last event (stop tracing).

As mentioned before, a pre-recorded file is used to transmit 31,726 events via the Target Tracer to the Device Gateway where they are forwarded to the Session Service. We would like to emphasize that we run all target tracer and thus emulated devices on a dedicated PC in order not to falsify the results of the scaling of the architecture. The target tracer emulation and the cloud backend were connected via 1 Gbit/s Ethernet connection.

Special attention is paid to load balancing for large systems. While the communication between the session service and the device gateway is ensured by Apache Kafka and its load balancing, a distribution of requests to the session service regarding the sessions is handled by Docker Swarm.
For our evaluation, different configurations are used, each of which is repeated ten times. Note that a separate target tracer is instantiated for each created session.

\begin{figure}
    \subfloat[Distribution of the processing time for a single session.]{
        \includegraphics[width=0.475\textwidth]{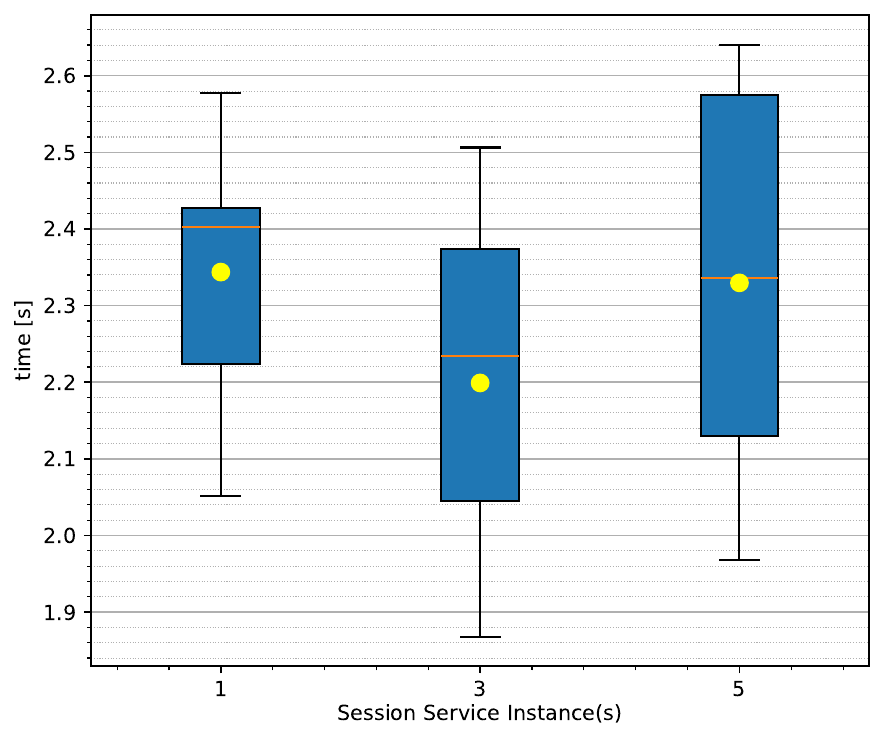}
        \label{subfig:proc_time_single_session}
    }
    \hfil
    \subfloat[Distribution of the processing time for 50 sessions.]{
        \includegraphics[width=0.475\textwidth]{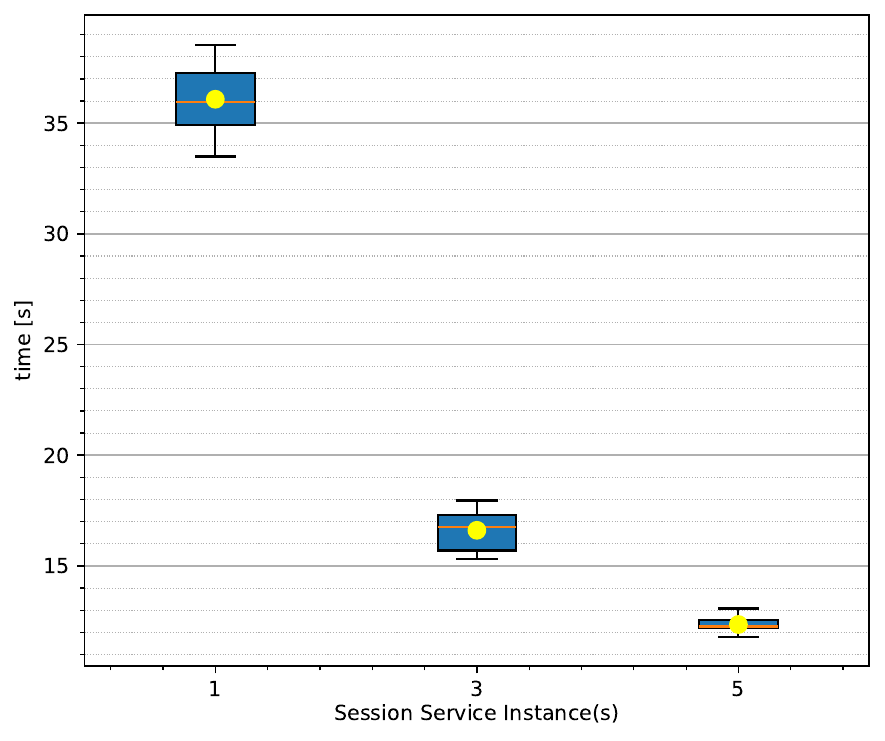}
        \label{subfig:proc_time_fifty_session}
    }
    
    \caption{Processing time results for the architecture.}
    \label{fig_sim}
    
\end{figure}

The results of the throughput measurement obtained can be seen in Table \ref{tab:evaluation_scaling_conf}, while we have shown the distribution of the processing time in Figures \ref{subfig:proc_time_single_session} and \ref{subfig:proc_time_fifty_session}. 

A single trace is processed for configuration 1 on an average of 2.3~s, whereby only a single session service is started. Therefore, the throughput results in 13,558 events per second.

In configuration no. 2, we virtually increase the number of simultaneously started sessions to 50, but only use a single service that manages all sessions alone. This increases the total processing time of a trace per session to an average of 36.1~s. Consequently, the throughput per session decreases to 879 events per second, while the total throughput increases to 43,966 events per second.

We are increasing the number of session services to three in order to evaluate the next two configurations. This reduces the processing time for a single session to 2.2~s or the throughput per session of 14,421 events per second. Since we only start a single Target Tracer, in this case only one session is created and is accordingly processed by a single session service, even if a total of three of these services have been started. We can also use this configuration to check the load balancing of Apache Kafka and Docker Swarm.

In configuration no. 4, we also increased the number of sessions. This allows us to test the scaling behavior. The results show that the average processing time increases to 16.6~s for 50 sessions, which corresponds to a throughput of 1,911 events per second. Thus, total throughput increases to 95,560 events per second, which is an increase of 117~\% compared to configuration no. 2 and thus shows that scaling through microservices works in our presented architecture.

We increase the number of sessions once again to a total of five. We get a slightly worse average processing time of 2.32~s in configuration no. 5 compared to configuration no. 3 with three Session Services. This reduces our throughput to 13,616 events per second if we start with only one tracing session.

In our last configuration no. 6, we again start 50 tracing sessions in parallel. These are now scaled to five session services, resulting in an average processing time of 12.35~s. The throughput per session increases to 2,569 events per second. Consequently, the total throughput increases to 128,445 events per second, which is almost three times as much as configuration no. 2 without scaling.

\section{Conclusion}\label{sec:conclusion}
This paper presents an architecture for cloud-based tracing of real-time embedded systems to ensure improved analysis and fault detection. The architecture consists of a target and a cloud component. We have shown that a cloud structure can bring advantages in terms of scaling such as memory and computational power to support a large number of microcontrollers generating traces simultaneously. 
We have shown how the presented architecture can be realized and implemented with its components, using different technologies and patterns like containerization and microservices. In an evaluation, we were able to show that the presented architecture enables scaling and thus provides a performance gain in the processing of software traces.

This architecture represents a cornerstone and provides a good basis for extensions. The following points are worth mentioning for future work: We used only software instrumentation for the previous instrumentation. The number of events could be increased by using non-intrusive tracing.
So far, the bidirectional connection between the cloud and the target tracer is unencrypted. An extension of the communication including encryption and certificates of the target tracers is conceivable.
Although the architecture offers scaling of the Session Service, the Device Gateway remains outside of the scope. The scaling of this application is hardly given in the current architecture. However, commercial solutions already exist for this service, such as AWS IoT Core.

Currently, each connected target system uses its own Target Tracer. Part of the future work is to investigate whether the efficiency of the Target Tracer can be increased by connecting more than one debugger per Target Tracer.

An extension towards observability of the architecture is also possible by supporting the Open Telemetry protocol. In this way, distributed tracing can be supported in the cloud infrastructure. Target systems that have a network connection can use software-based tracing to send generated events directly to the cloud. The target tracer component is no longer required for these target systems.

\bibliographystyle{IEEEtran}
\input{ecrts_rt-cloud_2025.bbl}

\end{document}